# Hyperfine Structure Constants of Sc I and Sc II with Fourier Transform Spectroscopy

Hala and G. Nave
National Institute of Standards and Technology, 100 Bureau Drive, Gaithersburg, MD 20899, USA; fnu.hala@nist.gov, hala.noman@gmail.com

## Abstract

We report new experimental hyperfine structure (HFS) constants of neutral and singly ionized scandium (Sc I and Sc II). We observed spectra of Sc–Ar and Sc–Ne hollow cathode discharges in the region 200–2500 nm (50,000–4000 cm$^{-1}$) using Fourier transform spectrometers. The measurements show significant HFS patterns in 1431 spectral lines fitted in our 12 spectra given in Table 1. These were fitted using the computer package Xgremlin to determine the magnetic dipole hyperfine interaction constant (A) for 185 levels in Sc I and 6 levels in Sc II, of which 80 Sc I levels had no previous measurements. The uncertainty in the HFS A constant is between $1 \times 10^{-4}$ and $5 \times 10^{-4}$ cm$^{-1}$.

## 1. Introduction

Scandium (Sc) is the lightest member of the Fe-group elements ($21 \leqslant Z \leqslant 28$), with atomic number 21, and it plays a vital role in astrophysical investigations because of its high stellar abundance and line-rich spectra. Knowledge of hyperfine structure (HFS) and isotopic shift (IS) measurement is important not only for theoretical atomic physics, but also for determining the elemental abundances in stellar atmospheres (Pickering 1996). Because of Doppler and rotational broadening, the HFS and IS are too small to be resolved in most astrophysical spectra, but they contribute to the overall profile of a spectral line (Abt 1952). Determining the elemental abundance from a spectral line often requires knowledge of HFS, to correct for saturation, and ignoring the underlying structure arising from HFS and IS can lead to errors as large as two to three orders of magnitude (Dylan et al. 2017). Because of this importance, there have been numerous publications on the HFS measurements of various levels in neutral and singly ionized scandium (Sc I and Sc II) using different experimental techniques. Radiofrequency (RF) techniques can measure magnetic dipole and electric dipole hyperfine interaction constants (A and B) with uncertainties in the kilohertz (1 kHz $\approx 3 \times 10^{-8}$ cm$^{-1}$) range, but RF techniques are limited only to the ground term. Kopfermann & Rasmussen (1934) used RF techniques to measure the first HFS constants of the two levels in the ground term 3d4s$^2$ $^2D_{3/2,5/2}$ of Sc I. These were later confirmed by Fricke et al. (1959) and Childs (1971) using atomic beam magnetic resonance (ABMR).

Single-frequency laser techniques can be used to measure HFS A and B constants with uncertainties typically in the MHz (1 MHz $\approx 3 \times 10^{-5}$ cm$^{-1}$) range, and have broader applicability. In Sc I, Ertmer & Hofer (1976) and Zeiske et al. (1976) combined single-frequency laser techniques with ABMR to study some low, metastable levels 3d$^2$($^3$F)4s $^4F_{3/2,5/2,7/2,9/2}$. Their precision and accuracy approached that of Childs (1971), and they demonstrated a technique with some applicability to levels above the ground term. Singh et al. (1991) combined a single-frequency laser with optogalvanic detection to measure HFS patterns for 3d$^2$(4p+4s) configurations of Sc I. Başar et al. (2004) combined a single-frequency laser with optogalvanic detection to measure HFS A constants for 3d4s4p and 3d4d5s levels in Sc I. Krzykowski & Stefańska (2008) used a single-frequency laser in laser-induced fluorescence and studied the even-parity levels in Sc I. Fourier transform (FT) spectroscopy instruments have the broadest applicability, but HFS A constants from FT spectroscopy measurements have typical uncertainties in the range of $10^{-5}$ to $10^{-4}$ cm$^{-1}$. FT spectroscopy instruments in the infrared (IR) were exploited in both early and more recent HFS studies on Sc I (Aboussaïd et al. 1996).

Although many low-lying levels now have accurately known HFS A constants, there are many higher levels in neutral and ionized Sc that still lack well-measured HFS constants. This work reports the first extensive studies of HFS measurements in a wide spectral range (200–2500 nm) using FT spectroscopy. The wavelength coverage of such a large spectral range allows us to measure the HFS constants for hundreds of levels by fitting the line profiles of thousands of lines. The HFS A and B constants for many levels can be derived from multiple lines in the UV-visible IR region.

The last and most extensive work on Sc I HFS is from Van Deelen (2017), who reported the HFS A constants for a total of 95 levels, of which 52 were completely new. Van Deelen (2017) confirmed the HFS A constants of the 3d$^2$($^1$D)4s $^2D_{3/2,5/2}$ levels, and found the Aboussaïd et al. (1996) values to be correct and rejected the Singh et al. (1991) values.

Before the analysis by Van Deelen (2017), a total of 49 HFS A constants were known, of which the measurements of Childs (1971), Zeiske et al. (1976), and Ertmer & Hofer (1976) are reported with high accuracy. The accuracy of the HFS A constants in those three publications (Childs 1971; Zeiske et al. 1976; Ertmer & Hofer 1976) is between 0.0003 % and 0.008%, whereas the HFS B constants have accuracies of 0.03% to 0.3%. Theoretical studies of HFS constants in both Sc I and Sc II also blossomed (e.g., Xi et al. 1991; Beck 1992; Chen 1994; Bieroń et al. 1995, 1997, 2002; Başar et al. 2004; Dembczyński et al. 2007; Őzturk et al. 2007) during the years of improved measurement techniques, but the best measurements are typically

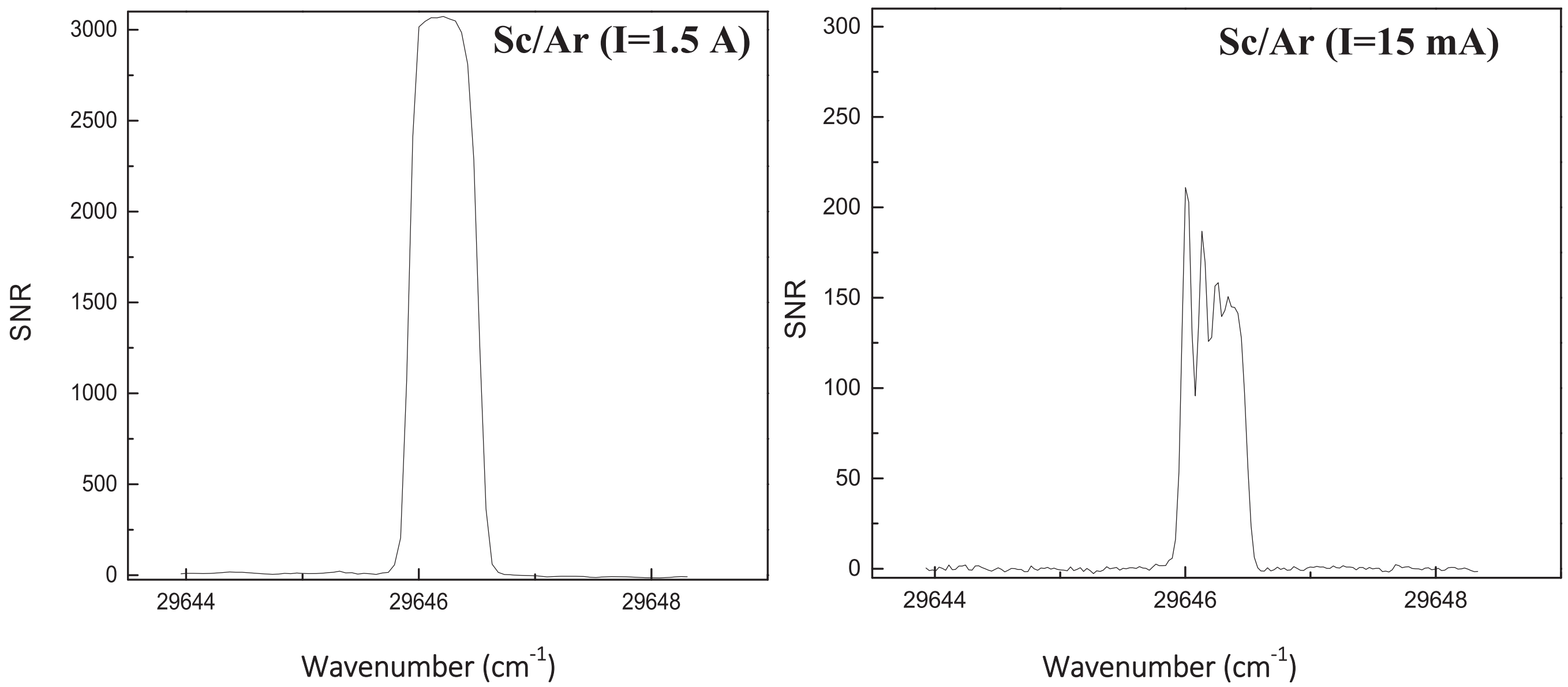

**Figure 1.** Portions of high-current (1.5 A; left) and low-current (15 mA; right) Sc/Ar spectra taken with water-cooled HCL (Danzmann et al. 1988) and commercial low-current HCL, respectively, showing strong self-absorption for a 29,646 $cm^{-1}$ line in the high-current spectrum.

preferred over theoretical values when measurements are available.

This research is carried out to derive new Sc abundances in the Sun, the mildly metal-poor (MP) red giant Arcturus, and the very MP halo main-sequence turnoff star HD 84937 (Lawler et al. 2019). Knowledge of HFS constants in Sc I and Sc II is essential in fitting the astrophysical spectra to synthesize the spectrum and the determination of elemental abundance. The goal of this study is to provide an extensive set of HFS constants for levels involved in the abundance determination of spectral synthesis. They include the HFS constants for the levels required for Lawler et al. (2019) to measure the Sc II abundance in MP stars.

Scandium consists of only one stable isotope, $^{45}$Sc, thereby removing any complications to line measurements due to blended isotopic structures. Due to the high nuclear spin (I) of 7/2 and a high magnetic dipole moment of +4.756 nuclear magnetons,[1] many transitions in Sc I and Sc II show wide structures (see Figure 1). The interaction of this magnetic moment with the electron field splits the energy levels into the smaller of the 2J+1 or 2I+1 components.

In the absence of perturbations, the energy of HFS level ($E_F$) is given by Emery (1996) as

$$E_F = E_J + \frac{1}{2}AK + B\frac{(3/4)K(K+1) - J(J+1)I(I+1)}{2I(2I-1)J(2J-1)}, \quad (1)$$

where $E_J$ is the energy of the fine structure level of quantum number $J$; $A$, and $B$ are the magnetic dipole and electric dipole hyperfine interaction constants, respectively; and $K$ is defined as

$$K = F(F+1) - J(J+1) - I(I+1), \quad (2)$$

where $F$ is the quantum number associated with the total angular momentum of the atom,

$$F = I + J;\, I + J - 1;\, ....;\, |I - J|. \quad (3)$$

[1] https://physics.nist.gov/PhysRefData/Handbook/Tables/scandiumtable1.htm

The selection rules that govern the hyperfine transitions are $\Delta F = 0, \pm 1$, but $F = 0 \leftrightarrow F' = 0$ are not allowed. In addition to this, there is a rule for intensity (Kuhn 1962) that says that within a hyperfine multiple, the ratio of the sums of the intensities of all transitions from two states with quantum numbers $F$ and $F'$ are in the ratio of their statistical weight $(2F+1)$:$(2F'+1)$. We employed this theory in fitting the HFS of Sc I and Sc II transitions using the Xgremlin Package (Nave et al. 2015). In most cases, the HFS components are not so regularly spaced, and the line profile can take many asymmetric forms.

## 2. Experiment

The spectra used for this study were taken on two National Institute of Standards and Technology (NIST) Fourier transform spectrometers (FTS). These spectra, listed in Table 1, were taken over a period of five years, mainly for wavelengths, energy levels, and branching fraction measurements, and are listed in Table 1. Spectral regions from 16,000 to 50,000 $cm^{-1}$ (625 to 200 nm) were taken with the NIST FT700 vacuum ultraviolet FTS (Griesmann et al. 1999). The region from 4000 to 40,000 $cm^{-1}$ (2500 to 250 nm) was recorded with the NIST 2 m FTS. A resolution of 0.01–0.03 $cm^{-1}$ was used, depending on the wavelength region, to resolve the Doppler-broadened line profiles of the transitions. The scandium was excited in high-current water-cooled hollow cathode lamps (HCLs; Danzmann et al. 1988), and commercial low-current HCLs operated at 9 to 15 mA. Commercial low-current HCLs run at low currents for long periods of time in a very stable fashion, and thus eliminate many optical depth problems (See Figure 1). These sealed HCLs are listed in Table 1 with a commercial supplier, including Hamamatsu (Hama.) and Varsal.

The high-current lamp used a graphite cathode lined with a solid scandium foil (99.99% pure) of thickness 0.2 mm, and operated at currents from 100 mA to 1.5 A with either argon (Ar) or neon (Ne) as buffer gas. The spectra taken with neon as a buffer gas enhanced the excitation of metal ions, and were useful for detecting the blends between Sc and Ar lines. Two

**Table 1**
Fourier Transform Spectra of Hollow Cathode Lamps Used in This Study

| Index Number | FTS | Spectra ID | Coadds[b] | Hollow Cathode Lamp[c] | Buffer Gas | Current (mA) | Useful Wavenumber Range[d] ($cm^{-1}$) | Resolution ($cm^{-1}$) |
|---|---|---|---|---|---|---|---|---|
| 1 | 2 m FTS | Sc051815.002 | 40 | Custom water-cooled | Ne | 1500 | 4000–20,000 | 0.015 |
| 2 | 2 m FTS | Sc051915.001 | 48 | Custom water-cooled | Ar | 1500 | 4000–19,000 | 0.010 |
| 3 | 2 m FTS | Sc120817.003 | 10 | Custom water-cooled | Ar | 200 | 9000–40,000 | 0.050 |
| 4 | FT700 | Sc092917.007[a] | 51 | Custom water-cooled | Ar | 100 | 16,000–50,000 | 0.030 |
| 5 | 2 m FTS | Sc050715b.005 | 128 | Sealed Commer. Varsal | Ar | 15 | 8000–30,000 | 0.030 |
| 6 | 2 m FTS | Sc050715a.005 | 128 | Sealed Commer. Varsal | Ar | 15 | 4100–12,000 | 0.030 |
| 7 | 2 m FTS | Sc050715.002 | 25 | Sealed Commer. Varsal | Ar | 9 | 9000–25,000 | 0.030 |
| 8 | 2 m FTS | Sc050715.001 | 25 | Sealed Commer. Varsal | Ar | 12 | 9000–25,000 | 0.030 |
| 9 | 2 m FTS | Sc050615.006 | 110 | Sealed Commer. Varsal | Ar | 15 | 9000–25,000 | 0.030 |
| 10 | FT700 | Sc050615.004 | 128 | Sealed Commer. Hama. | Ar | 15 | 16,000–50,000 | 0.030 |
| 11 | FT700 | Sc050515a.003 | 112 | Sealed Commer. Hama. | Ar | 15 | 16,000–50,000 | 0.030 |
| 12 | FT700 | Sc092917.004[a] | 4 | Custom water-cooled | Ar | 100 | 16,000–50,000 | 0.030 |

**Notes.** Any commercial equipment, instruments, or materials are identified in this paper to foster understanding. Such identification does not imply recommendation or endorsement by the National Institute of Standards and Technology, nor does it imply that the materials or equipment identified are necessarily the best available for the purpose.
[a] Spectra were taken with germanium (Ge) pieces in the cathode.
[b] Coherent additions of interferograms (coadds) determine the total data collection times and S/N levels in the spectra.
[c] Hollow cathode lamps include sealed commercial (Commer.) lamps that run at low currents for long periods of time in a very stable fashion, and thus eliminate many optical depth problems. These sealed HCLs are listed with a commercial supplier, including Hamamatsu (Hama.) A custom demountable lamp that is water-cooled and able to run at high current is also listed.
[d] The useful wavenumber range is listed here, which may be smaller than measured.

spectra (Sc092917.007 and Sc092917.004) were taken with germanium (Ge) pieces in the cathode to provide wavelength standards, and were useful for the HFS study of highly excited levels. The high-current spectra are more useful for singly ionized scandium (Sc II) analysis due to their enhanced signal-to-noise ratios (S/Ns). A comparison of spectra taken at various currents enabled us to identify HFS profiles affected by self-absorption, as shown in Figure 1.

## 3. Hyperfine Structure Analysis of Sc I

For the HFS analysis, the interferograms from the FTS were transformed and phase-corrected with the program Xgremlin (Nave et al. 2015), which is also used to determine the HFS A and B constants. The Xgremlin program (Nave et al. 2015) incorporates the HFS fitting routines of Pulliam (1977) to fit up to three transitions to a spectral feature. Eight parameters may be either fixed or allowed to vary: HFS A and B constants of the upper and lower levels, the center of gravity wavenumber of the line, the peak intensity of the strongest HFS component, the FWHM of the HFS components, and the ratio of the Lorentzian width to the total width. Initial values of these parameters for each transition are either estimated or obtained from previous measurements. The program then refines these values with an iterative least-squares fit.

As the starting point of the analysis, we used the highly accurate HFS A and B constants for the 3d4s$^2$ $^2$D$_{3/2,5/2}$ levels at 0 cm$^{-1}$ and 168.34 cm$^{-1}$ from Childs (1971), and the four metastable levels 3d$^2$($^3$F)4s $^4$F$_{3/2,5/2,7/2,9/2}$ at 11,519.99 cm$^{-1}$, 11,557.69 cm$^{-1}$, 11,610.28 cm$^{-1}$, and 11,677.38 cm$^{-1}$ by Zeiske et al. (1976) and Ertmer & Hofer (1976), respectively. The HFS A and B constants for these levels are known very precisely, with an uncertainty in the HFS A constants of 0.0003% to 0.008%, and an uncertainty in the HFS B constants of 0.03% to 0.3%. We used these values of HFS constants to measure the HFS A and B constants for the odd-parity levels (3d4s4p, 4s$^2$4p, and 3d$^2$4p) that connect to the 3d4s$^2$ $^2$D and 3d$^2$($^3$F)4s $^4$F terms with partially resolved HFS patterns. We kept the HFS A and B constants for even-parity levels as a fixed parameter in the fit. As a result, we measured the HFS A constants for 26 levels belonging to the odd-parity levels of the 3d4s4p, 3d$^2$4p, and 4s$^2$4p configurations. These were, in turn, used to determine the HFS constants of the even-parity levels belonging to the 3d4s5s, 3d4s4d, 3d$^2$4s, and 3d$^3$ configurations, and continued this process to determine the HFS constants for all other levels.

All of the measured HFS constants for even parity are given in Table 2, and those for odd parity are given in Table 3. The first three columns give the configuration, term designation, and $J$-value of the energy level taken from Kramida et al (2020). Column (4) gives the energy of the level taken from Ahmed & Verges (1977). Column (5) gives our experimental HFS A constant, with the estimated uncertainty of the last digits (one standard deviation) in parenthesis. Column (6) gives our experimental HFS B constant, with the estimated uncertainty in parenthesis. Column (7) gives the total number of spectral lines used to derive the HFS A and B constants. Column (8) has information about the spectra (listed in Table 1) used for fitting the spectral line in Xgremlin (Nave et al. 2015). The lowest-uncertainty HFS A and B constants from the literature are given in columns 9 and 10, with the publication references in column (11). The uncertainties in HFS constants (see columns 9 and 10 of Tables 2 and 3) measured by Van Deelen (2017) are unrealistic, and therefore we did not use them in this work.

Singh et al. (1991) and Aboussaïd et al. (1996) both measured the HFS A constant for 3d$^2$($^1$D)4s $^2$D$_{3/2,5/2}$, but they did not agree with each other. Later, Başar et al. (2004) pointed them out to be ambiguous values. In the present work, we confirmed the HFS constant of Aboussaïd et al. (1996) and rejected the Singh et al. (1991) value. Three other levels 3d$^2$($^3$F)4p $^2$F$^*_{7/2}$ (33,278.40 cm$^{-1}$), 3d$^2$($^3$F)4p $^2$D$^*_{3/2}$ (33,614.88 cm$^{-1}$), and 3d$^2$($^3$F)4p $^2$D$^*_{5/2}$ (33,707.06 cm$^{-1}$) studied by Singh et al. (1991) are marked as ambiguous in Başar et al. (2004), and therefore need further confirmation. In this work, we measured the HFS A constants for these three levels using at least three spectral lines in multiple spectra, and rejected the values of the HFS A constants for these three levels ($E$ = 33,278.40 cm$^{-1}$, 33,614.88 cm$^{-1}$, and 33,707.06 cm$^{-1}$) by Singh et al. (1991). We confirmed that the values of the HFS A constants from Başar et al. (2004; $E$ = 33,614.88 cm$^{-1}$ and 33,707.06 cm$^{-1}$) are correct.

The levels at 24,656.72 cm$^{-1}$ and 24,656.88 cm$^{-1}$ (3d4s($^1$D)4p $^2$P$^*_{1/2,3/2}$) are very close to each other, and need a special mention. First, we studied the 24,656.88 cm$^{-1}$ (3d4s($^1$D)4p $^2$P$^*_{3/2}$) decaying to the 17,012.76 cm$^{-1}$ (3d$^2$($^1$D)4s $^2$D$_{5/2}$) level at wavenumber 7644.12 cm$^{-1}$, as this line is unblended. We determined the HFS constants for the 24,656.88 cm$^{-1}$ level using a single transition with a well-resolved HFS profile at 7644.12 cm$^{-1}$, giving A = (0.42 ± 0.03) mK and B = (−6.36 ± 0.42) mK. The 24,656.72 cm$^{-1}$ (3d4s($^1$D)4p $^2$P$^*_{1/2}$) level is connected to the 17,025.14 cm$^{-1}$ (3d$^2$($^1$D)4s $^2$D$_{3/2}$) level by a transition at 7631.58 cm$^{-1}$. This line is mixed with the 3d$^2$($^1$D)4s $^2$D$_{3/2}$ ↔ 3d4s($^1$D)4p $^2$P$^*_{3/2}$ transition at 7631.74 cm$^{-1}$, as shown in Figure 2 by the red dotted line. We determined the value of the HFS A constant for the 24,656.72 (3d4s($^1$D)4p $^2$P$^*_{1/2}$) level by fitting the two HFS profiles simultaneously in Xgremlin (Nave et al. 2015), and fixing the HFS constants for the 3d4s($^1$D)4p $^2$P$^*_{3/2}$ (24,656.88 cm$^{-1}$) and 17,025.14 cm$^{-1}$ levels to derive the HFS constant for the 24,656.72 cm$^{-1}$ (3d4s($^1$D)4p $^2$P$^*_{1/2}$) level. The observed and fitted profiles are shown in Figure 2.

In this work, we measured the HFS constants for 176 levels of Sc I, of which 80 are new and 8 are revised HFS A constants. Out of 207 levels belonging to the {3d(4s$^2$+4p$^2$), 3d$^2$(4s+5s+4p+4d), 4s$^2$(4p+4d), 3d4s(5s+6s+4p+5p+4d), and 3d$^3$} configurations in Sc I, a total of 185 levels now have known HFS A constants, and are given in Tables 2 and 3. Table 4 includes the positions and relative strengths of individual HFS components for 13,494 lines of Sc I.

## 4. Hyperfine Structure Analysis of Sc II

Work on the HFS constant of Sc II is less common. The first work on HFS in Sc II was reported back in 1982 by Arnesen et al. (1982), using the collinear laser ion beam technique. Arnesen et al. (1982) reported the HFS A and B constants for eight levels belonging to 3d$^2$ ($^1$D, $^3$P$_{1,2}$) and 3d4p($^1$D$^*_2$, $^3$D$^*_{1,2,3}$, $^3$P$^*_1$). Later, Young et al. (1988) studied the HFS constants using the collinear laser ion beam method for 3d$^2$ and 3d4p configurations. They confirmed the HFS A and B constants by Arnesen et al. (1982), and reported five new HFS A and B constants belonging to these two configurations (3d$^2$ $^1$G$_4$, 3d4p $^3$F$^*_{2,3}$, 3d4p $^3$P$^*_2$, 3d4p $^1$F$^*_3$). Mansour et al. (1989a) used a collinear fast-beam experiment, and Mansour et al. (1989b) used a laser-RF double resonance experiment to measure the HFS A and B constants for levels of 3d$^2$ and 3d4p

**Table 2**
New and Remeasured Hyperfine Structure Constants for Even Levels in Sc I

| Level Designation[a] | | | Energy[a] ($cm^{-1}$) | This Work | | | | Previous Experimental Results[e] | | |
|---|---|---|---|---|---|---|---|---|---|---|
| Configuration | Term | J | | A(ΔA)[b] (mK)[c] | B(ΔB)[b] (mK)[c] | No. of Lines[d] | Spectra Used | A(ΔA)$^{prev}$ (mK)[c] | B(ΔB)$^{prev}$ (mK)[c] | Reference |
| $3d4s^2$ | $^2D$ | 3/2 | 0 | … | … | … | … | **8.99149(3)** | **0.8793(3)** | Childs (1971) |
| $3d4s^2$ | $^2D$ | 5/2 | 168.34 | … | … | … | … | **3.63695(3)** | **−1.2466(5)** | Childs (1971) |
| $3d^2(^3F)4s$ | $^4F$ | 3/2 | 11,519.99 | … | … | … | … | **−5.2874(1)** | **−0.1731(7)** | Ertmer & Hofer (1976) |
| $3d^2(^3F)4s$ | $^4F$ | 5/2 | 11,557.69 | … | … | … | … | **5.13806(7)** | **−0.215(1)** | Zeiske et al. (1976) |
| $3d^2(^3F)4s$ | $^4F$ | 7/2 | 11,610.28 | … | … | … | … | **8.33950(7)** | **−0.303(1)** | Zeiske et al. (1976) |
| $3d^2(^3F)4s$ | $^4F$ | 9/2 | 11,677.38 | … | … | … | … | **9.5388(3)** | **−0.516(0.023)** | Ertmer & Hofer (1976) |
| $3d^2(^3F)4s$ | $^2F$ | 5/2 | 14,926.07 | **9.73(19)** | … | 2(3) | 1,2 | 9.19(0.011) | −0.21(0.00021) | Van Deelen (2017) |
| $3d^2(^3F)4s$ | $^2F$ | 7/2 | 15,041.92 | **−0.60(4)** | … | 1(2) | 1,2 | −0.54(0.018) | … | Van Deelen (2017) |
| $3d^2(^1D)4s$ | $^2D$ | 5/2 | 17,012.76 | **18.43(2)** | **2.97(38)** | 3(5) | 1,2 | 18.38(33) | … | Aboussaïd et al. (1996) |
| $3d^2(^1D)4s$ | $^2D$ | 3/2 | 17,025.14 | **−9.62(3)** | **−0.56(28)** | 1(2) | 1,2 | −9.51(83) | … | Aboussaïd et al. (1996) |
| $3d^2(^3P)4s$ | $^4P$ | 1/2 | 17,226.04 | **36.24(16)** | … | 3(14) | 1–5; 7–12 | … | … | … |
| $3d^2(^3P)4s$ | $^4P$ | 3/2 | 17,255.07 | **16.44(17)** | … | 6(29) | 1–12 | … | … | … |
| $3d^2(^3P)4s$ | $^4P$ | 5/2 | 17,307.08 | **15.74(6)** | … | 8(29) | 1–12 | 14.51(73) | … | Krzykowski & Stefańska (2008) |
| $3d^2(^1G)4s$ | $^2G$ | 9/2 | 20,236.86 | **11.87(13)** | **10.13(1.04)** | 1(2) | 1,2 | 11.70(0.0036) | … | Van Deelen (2017) |
| $3d^2(^1G)4s$ | $^2G$ | 7/2 | 20,239.66 | **−3.65(11)** | … | 1(2) | 1,2 | −3.83(0.0077) | … | Van Deelen (2017) |
| $3d^2(^3P)4s$ | $^2P$ | 1/2 | 20,681.43 | **11.98(17)** | … | 2(4) | 1,2 | 11.71(17) | … | Krzykowski & Stefańska (2008) |
| $3d^2(^3P)4s$ | $^2P$ | 3/2 | 20,719.86 | **−6.98(6)** | **0.77(69)** | 3(5) | 1,2 | −7.11(33) | … | Basar et al.(2004) |
| $3d^2(^1S)4s$ | $^2S$ | 1/2 | 26,936.98 | **82.11(1)** | … | 4(6) | 1,2 | 82.08(0.00089) | … | Van Deelen (2017) |
| $3d^3$ | $^4F$ | 3/2 | 33,763.53 | **9.36(3)** | … | 1(1) | 1–5,7–9,12 | … | … | … |
| $3d^3$ | $^4F$ | 5/2 | 33,798.64 | **3.15(3)** | … | 4(17) | 1,2,4,5,9,12 | … | … | … |
| $3d^3$ | $^4F$ | 7/2 | 33,846.59 | **1.15(1)** | … | 5(33) | 1,2,4–12 | 1.00(0.0084) | … | Van Deelen (2017) |
| $3d^3$ | $^4F$ | 9/2 | 33,906.38 | **0.29(1)**[f] | … | 3(24) | … | 1.44(0.0023) | … | Van Deelen (2017) |
| $3d4s(^3D)5s$ | $^4D$ | 1/2 | 34,390.25 | **−24.67(8)** | … | 3(11) | 1–10,12 | −24.62(10) | … | Başar et al. (2004) |
| $3d4s(^3D)5s$ | $^4D$ | 3/2 | 34,422.83 | **15.03(14)** | … | 3(13) | 1–4,12 | 14.94(07) | … | Krzykowski & Stefańska (2008) |
| $3d4s(^3D)5s$ | $^4D$ | 5/2 | 34,480.00 | **19.85(2)** | … | 4(21) | 1–5,9,12 | 19.75(07) | … | Krzykowski & Stefańska (2008) |
| $3d4s(^3D)5s$ | $^4D$ | 7/2 | 34,567.19 | **20.14(4)** | … | 3(16) | 1–5,8,9 | 20.58(20) | … | Krzykowski & Stefańska (2008) |
| $3d4s(^3D)5s$ | $^2D$ | 3/2 | 35,671.04 | **−5.56(19)** | … | 3(6) | 1,4,5,8,9 | −5.54(20) | … | Krzykowski & Stefańska (2008) |
| $3d4s(^3D)5s$ | $^2D$ | 5/2 | 35,745.62 | **20.06(10)** | … | 2(13) | 1,3–5,7–10 | 20.21(17) | … | Krzykowski & Stefańska (2008) |
| $3d^3$ | $^2D2$ | 3/2 | 36,276.63 | **1.59(33)** | … | 1(2) | 4,12 | 1.43(3) | … | Krzykowski & Stefańska (2008) |
| $3d^3$ | $^2D2$ | 5/2 | 36,330.59 | **5.60(5)** | … | 4(9) | 1–5,7,8 | 5.70(3) | … | Krzykowski & Stefańska (2008) |
| $3d^3$ | $^4P$ | 1/2 | 36,492.64 | **−11.25(18)** | … | 2(6) | 1,2,4,9,12 | −11.17(20) | … | Krzykowski & Stefańska (2008) |
| $3d^3$ | $^4P$ | 3/2 | 36,515.76 | **−0.91(7)** | … | 5(13) | 1,2,4,5,9,12 | … | … | … |
| $3d^3$ | $^4P$ | 5/2 | 36,572.77 | **−0.67(2)** | … | 4(17) | 1–5,7–12 | −0.83(7) | … | Krzykowski & Stefańska (2008) |
| $3d^3$ | $^2G$ | 7/2 | 36,977.51 | **4.95(17)** | … | 2(3) | 1,2 | 5.14(0.0042) | … | Van Deelen (2017) |
| $3d^3$ | $^2G$ | 9/2 | 37,054.51 | **1.68(10)** | … | 2(3) | 1,2 | 1.75(0.0032) | … | Van Deelen (2017) |
| $3d^3$ | $^2P$ | 1/2 | 37,085.84 | **1.67(15)** | … | 1(3) | 4,9,12 | … | … | … |
| $3d^3$ | $^2P$ | 3/2 | 37,148.22 | **2.98(8)** | … | 3(10) | 1,2,4,5,8,9,11,12 | … | … | … |
| $3d4s(^1D)5s$ | $^2D$ | 3/2 | 37,780.87 | **10.16(3)** | … | 2(6) | 1,2,4,5,9,12 | 9.77(10) | … | Krzykowski & Stefańska (2008) |
| $3d4s(^1D)5s$ | $^2D$ | 5/2 | 37,855.61 | **0.10(7)** | … | 2(10) | 1–5,8,9,12 | −0.51(0.004) | … | Van Deelen (2017) |
| $3d4s(^3D)4d$ | $^2F$ | 5/2 | 38,871.65 | **−3.53(11)** | … | 1(3) | 4,9,12 | −3.47(0.0084) | … | Van Deelen (2017) |
| $3d4s(^3D)4d$ | $^2F$ | 7/2 | 38,959.16 | **10.92(8)** | … | 1(2) | 4,12 | 10.89(1) | … | Van Deelen (2017) |
| $3d^3$ | $^2H$ | 9/2 | 39,164.11 | … | … | … | … | … | … | … |
| $3d^3$ | $^2H$ | 11/2 | 39,225.33 | **2.00(19)** | … | 1(2) | 1,2 | 1.99(0.0019) | … | Van Deelen (2017) |
| $3d4s(^3D)4d$ | $^4D$ | 1/2 | 39,701.44 | **−29.56(26)** | … | 3(7) | 4,9,12 | … | … | … |
| $3d4s(^3D)4d$ | $^4D$ | 3/2 | 39,721.79 | **9.62(18)** | … | 3(6) | 4,12 | … | … | … |

**Table 2**
(Continued)

| Level Designation[a] | | | Energy[a] (cm$^{-1}$) | This Work | | | | Previous Experimental Results[e] | | |
|---|---|---|---|---|---|---|---|---|---|---|
| Configuration | Term | J | | A(ΔA)[b] (mK)[c] | B(ΔB)[b] (mK)[c] | No. of Lines[d] | Spectra Used | A(ΔA)$^{prev}$ (mK)[c] | B(ΔB)$^{prev}$ (mK)[c] | Reference |
| 3d4s($^3$D)4d | $^4$D | 5/2 | 39,755.02 | **14.69(6)** | ⋯ | 5(13) | 4,9–12 | ⋯ | ⋯ | ⋯ |
| 3d4s($^3$D)4d | $^4$D | 7/2 | 39,799.99 | **16.78(2)** | ⋯ | 2(7) | 4,5,9,10,12 | ⋯ | ⋯ | ⋯ |
| 3d4s($^3$D)4d | $^4$G | 5/2 | 39,861.37 | **−9.71(6)** | ⋯ | 3(5) | 4,9,12 | ⋯ | ⋯ | ⋯ |
| 3d4s($^3$D)4d | $^4$G | 7/2 | 39,902.75 | **3.38(7)** | ⋯ | 2(4) | 4,12 | ⋯ | ⋯ | ⋯ |
| 3d4s($^3$D)4d | $^4$G | 9/2 | 39,957.79 | **8.71(22)** | ⋯ | 1(2) | 4,12 | ⋯ | ⋯ | ⋯ |
| 3d4s(3D)4d | $^4$G | 11/2 | 40,028.38 | ⋯ | ⋯ | ⋯ | ⋯ | ⋯ | ⋯ | ⋯ |
| 3d4s($^3$D)4d | $^2$P | 3/2 | 40,063.88 | **24.37(11)** | ⋯ | 3(5) | 1,2,4,9 | ⋯ | ⋯ | ⋯ |
| 3d4s($^3$D)4d | $^2$P | 1/2 | 40,070.30 | **−25.26(65)** | ⋯ | 2(5) | 1,2,4,9,12 | −24.55(0.049) | ⋯ | Van Deelen (2017) |
| 3d4s($^3$D)4d | $^2$D | 3/2 | 40,257.52 | **16.02(12)** | ⋯ | 3(9) | 1,3,4,9,12 | ⋯ | ⋯ | ⋯ |
| 3d4s($^3$D)4d | $^2$D | 5/2 | 40,334.31 | **5.27(0.19)** | ⋯ | 3(5) | 2,4,12 | ⋯ | ⋯ | ⋯ |
| 3d4s($^3$D)4d | $^4$S | 3/2 | 40,282.16 | **23.59(7)** | ⋯ | 4(11) | 4,5,9,10,12 | ⋯ | ⋯ | ⋯ |
| 3d4s($^3$D)4d | $^2$G | 7/2 | 40,418.55 | **3.79(12)** | ⋯ | 5(8) | 1,2,4 | 3.92(0.0035) | ⋯ | Van Deelen (2017) |
| 3d4s($^3$D)4d | $^2$G | 9/2 | 40,562.06 | **3.10(84)** | ⋯ | 3(6) | 1,2 | 3.77(0.011) | ⋯ | Van Deelen (2017) |
| 3d4s($^3$D)4d | $^4$F | 3/2 | 40,521.27 | **−7.18(32)** | ⋯ | 2(4) | 1,2,4,12 | ⋯ | ⋯ | ⋯ |
| 3d4s($^3$D)4d | $^4$F | 5/2 | 40,554.99 | **6.28(15)** | ⋯ | 2(3) | 4,12 | ⋯ | ⋯ | ⋯ |
| 3d4s($^3$D)4d | $^4$F | 7/2 | 40,603.95 | **9.32(5)** | ⋯ | 5(8) | 1,2,4,12 | ⋯ | ⋯ | ⋯ |
| 3d4s($^3$D)4d | $^4$F | 9/2 | 40,670.87 | **10.01(11)** | ⋯ | 2(4) | 1,2 | ⋯ | ⋯ | ⋯ |
| 3d$^3$ | $^2$F | 5/2 | 40,802.76 | **2.23(5)** | ⋯ | 2(7) | 3,4,9–12 | ⋯ | ⋯ | ⋯ |
| 3d$^3$ | $^2$F | 7/2 | 40,825.78 | **3.14(6)** | ⋯ | 1(6) | 3,4,9–12 | 2.65(0.019) | ⋯ | Van Deelen (2017) |
| 3d4s($^3$D)4d | $^4$P | 1/2 | 41,446.85 | **55.99(1.03)** | ⋯ | 2(4) | 4,10,12 | ⋯ | ⋯ | ⋯ |
| 3d4s($^3$D)4d | $^4$P | 3/2 | 41,474.87 | **25.89(46)** | ⋯ | 3(6) | 4,9,12 | 26.49(37) | ⋯ | Krzykowski & Stefańska (2008) |
| 3d4s($^3$D)4d | $^4$P | 5/2 | 41,505.60 | **22.09(7)** | ⋯ | 2(8) | 4,9–12 | 22.25(30) | ⋯ | Krzykowski & Stefańska (2008) |
| 3d$^2$($^3$F)5s | $^4$F | 3/2 | 41,921.89 | **−0.06(5)**[f] | ⋯ | 2(4) | 1,2 | 1.66(13) | ⋯ | Van Deelen (2017) |
| 3d$^2$($^3$F)5s | $^4$F | 5/2 | 41,960.97 | **5.78(87)**[f] | ⋯ | 2(2) | 4 | 3.30(0.028) | ⋯ | Van Deelen (2017) |
| 3d$^2$($^3$F)5s | $^4$F | 7/2 | 42,015.58 | **−2.85(4)** | ⋯ | 1(2) | 1,2 | ⋯ | ⋯ | ⋯ |
| 3d$^2$($^3$F)5s | $^4$F | 9/2 | 42,085.18 | **6.46(1)** | ⋯ | 1(2) | 1,2 | ⋯ | ⋯ | ⋯ |
| 3d4s($^1$D)4d | $^2$F | 5/2 | 42,149.66 | **2.99(6)** | ⋯ | 2(4) | 1,2 | ⋯ | ⋯ | ⋯ |
| 3d4s($^1$D)4d | $^2$F | 7/2 | 42,198.84 | **4.84(2)**[f] | ⋯ | 2(8) | 4,5,8–12 | 1.14(0.0018) | ⋯ | Van Deelen (2017) |
| 3d4s($^1$D)4d | $^2$D | 5/2 | 42,445.55 | **7.83(11)** | ⋯ | 2(4) | 4,12 | 8.04(34) | ⋯ | Krzykowski & Stefańska (2008) |
| 3d4s($^1$D)4d | $^2$D | 3/2 | 42,466.39 | ⋯ | ⋯ | ⋯ | ⋯ | −0.20(57) | ⋯ | Krzykowski & Stefańska (2008) |
| 3d4s($^3$D)4d | $^2$S | 1/2 | 42,877.65 | ⋯ | ⋯ | ⋯ | ⋯ | ⋯ | ⋯ | ⋯ |
| 3d$^3$ | $^2$D1 | 5/2 | 42,917.83 | **13.34(13)** | ⋯ | 2(8) | 1,2,4,9,10,12 | ⋯ | ⋯ | ⋯ |
| 3d$^3$ | $^2$D2 | 3/2 | 42,937.50 | **−5.39(1.24)** | ⋯ | 2(3) | 1,4 | ⋯ | ⋯ | ⋯ |
| 3d4s($^1$D)4d | $^2$G | 9/2 | 42,942.51 | ⋯ | ⋯ | ⋯ | ⋯ | ⋯ | ⋯ | ⋯ |
| 3d4s($^1$D)4d | $^2$G | 7/2 | 42,969.78 | **−5.85(46)** | ⋯ | 1(2) | 4,12 | −4.97(13) | ⋯ | Krzykowski & Stefańska (2008) |
| 3d4s($^1$D)4d | $^2$P | 1/2 | 43,429.68 | **7.49(1.10)** | ⋯ | 1(1) | 12 | ⋯ | ⋯ | ⋯ |
| 3d4s($^1$D)4d | $^2$P | 3/2 | 43,435.40 | **−1.97(52)** | ⋯ | 2(3) | 4,12 | ⋯ | ⋯ | ⋯ |
| 4s$^2$4d | $^2$D | 3/2 | 43,597.16 | **10.55(29)** | ⋯ | 1(2) | 4,12 | ⋯ | ⋯ | ⋯ |
| 4s$^2$4d | $^2$D | 5/2 | 43,658.53 | **0.05(25)** | ⋯ | 2(4) | 4,12 | ⋯ | ⋯ | ⋯ |
| 3d4s($^3$D)6s | $^4$D | 1/2 | 43,809.76 | ⋯ | ⋯ | ⋯ | ⋯ | ⋯ | ⋯ | ⋯ |
| 3d4s($^3$D)6s | $^4$D | 3/2 | 43,814.47 | **−11.04(50)** | ⋯ | 1(2) | 1,4 | ⋯ | ⋯ | ⋯ |
| 3d4s($^3$D)6s | $^4$D | 5/2 | 43,898.31 | ⋯ | ⋯ | ⋯ | ⋯ | ⋯ | ⋯ | ⋯ |
| 3d4s($^3$D)6s | $^4$D | 7/2 | 43,988.20 | ⋯ | ⋯ | ⋯ | ⋯ | ⋯ | ⋯ | ⋯ |
| 3d($^2$D)4p$^2$($^3$P) | $^4$P | 1/2 | 44,030.34 | **48.18(1.90)** | ⋯ | 1(2) | 4,12 | ⋯ | ⋯ | ⋯ |

**Table 2**
(Continued)

| Level Designation[a] | | | Energy[a] ($cm^{-1}$) | This Work | | | | Previous Experimental Results[e] | | |
|---|---|---|---|---|---|---|---|---|---|---|
| Configuration | Term | J | | A(ΔA)[b] (mK)[c] | B(ΔB)[b] (mK)[c] | No. of Lines[d] | Spectra Used | $A(\Delta A)^{prev}$ (mK)[c] | $B(\Delta B)^{prev}$ (mK)[c] | Reference |
| $3d(^2D)4p^2(^3P)$ | $^4P$ | 3/2 | 44,107.25 | **23.22**(**24**) | ⋯ | 3(5) | 4,12 | ⋯ | ⋯ | ⋯ |
| $3d(^2D)4p^2(^3P)$ | $^4P$ | 5/2 | 44,238.23 | **21.49**(**16**) | ⋯ | 2(4) | 1,12 | ⋯ | ⋯ | ⋯ |
| $3d(^2D)4p^2(^3P)$ | $^2P$ | 3/2 | 44,594.97 | **2.30**(**32**) | ⋯ | 2(3) | 4,12 | ⋯ | ⋯ | ⋯ |
| $3d(^2D)4p^2(^3P)$ | $^2P$ | 1/2 | 44,690.65 | **4.82**(**99**) | ⋯ | 1(2) | 4,12 | ⋯ | ⋯ | ⋯ |
| $3d(^2D)4p^2(^3P)$ | $^4F$ | 3/2 | 44,823.21 | **11.10**(**22**) | ⋯ | 2(5) | 4,10,12 | ⋯ | ⋯ | ⋯ |
| $3d(^2D)4p^2(^3P)$ | $^4F$ | 5/2 | 44,909.55 | **3.46**(**27**) | ⋯ | 2(3) | 4,12 | ⋯ | ⋯ | ⋯ |
| $3d(^2D)4p^2(^3P)$ | $^4F$ | 7/2 | 45,016.43 | **4.10**(**11**) | ⋯ | 2(5) | 4,10–12 | ⋯ | ⋯ | ⋯ |
| $3d(^2D)4p^2(^3P)$ | $^4F$ | 9/2 | 45,125.73 | **2.71**(**5**) | ⋯ | 3(11) | 3,4,10–12 | ⋯ | ⋯ | ⋯ |
| $3d(^2D)4p^2(^3P)$ | $^2F$ | 5/2 | 44,838.56 | **3.01**(**58**) | ⋯ | 2(1) | 4 | ⋯ | ⋯ | ⋯ |
| $3d(^2D)4p^2(^3P)$ | $^2F$ | 7/2 | 44,941.81 | **7.40** (**11**) | ⋯ | 4(7) | 4,12 | ⋯ | ⋯ | ⋯ |
| $3d4s(^1D)4d$ | $^2S$ | 1/2 | 45,514.98 | **34.91**(**1.78**) | ⋯ | 1(2) | 4,12 | ⋯ | ⋯ | ⋯ |
| $3d(^2D)4p^2(^3P)$ | $^4D$ | 1/2 | 45,574.64 | ⋯ | ⋯ | ⋯ | ⋯ | ⋯ | ⋯ | ⋯ |
| $3d(^2D)4p^2(^3P)$ | $^4D$ | 3/2 | 45,605.80 | ⋯ | ⋯ | ⋯ | ⋯ | ⋯ | ⋯ | ⋯ |
| $3d(^2D)4p^2(^3P)$ | $^4D$ | 5/2 | 45,659.09 | **6.08**(**23**) | ⋯ | 5(6) | 1,4,12 | ⋯ | ⋯ | ⋯ |
| $3d^2(^3F)4d$ | $^4G$ | 5/2 | 45,715.79 | ⋯ | ⋯ | ⋯ | ⋯ | ⋯ | ⋯ | ⋯ |
| $3d(^2D)4p^2(^3P)$ | $^4D$ | 7/2 | 45,737.17 | **8.88**(**8**) | ⋯ | 4(7) | 1,4,11,12 | ⋯ | ⋯ | ⋯ |
| $3d^2(^3F)4d$ | $^4G$ | 7/2 | 45,752.28 | ⋯ | ⋯ | ⋯ | ⋯ | ⋯ | ⋯ | ⋯ |
| $3d^2(^3F)4d$ | $^4G$ | 9/2 | 45,804.10 | ⋯ | ⋯ | ⋯ | ⋯ | ⋯ | ⋯ | ⋯ |
| $3d^2(^3F)4d$ | $^4G$ | 11/2 | 45,870.92 | **0.72**(**23**) | ⋯ | 1(2) | 4,12 | ⋯ | ⋯ | ⋯ |
| $3d^2(^3F)4d$ | $^4H$ | 7/2 | 45,878.06 | ⋯ | ⋯ | ⋯ | ⋯ | ⋯ | ⋯ | ⋯ |
| $3d^2(^3F)4d$ | $^4H$ | 9/2 | 45,925.09 | ⋯ | ⋯ | ⋯ | ⋯ | ⋯ | ⋯ | ⋯ |
| $3d^2(^3F)4d$ | $^4D$ | 1/2 | 45,927.81 | **−17.10**(**2.32**) | ⋯ | 2(3) | 1,2,4 | ⋯ | ⋯ | ⋯ |
| $3d^2(^3F)4d$ | $^4D$ | 3/2 | 45,947.35 | ⋯ | ⋯ | ⋯ | ⋯ | ⋯ | ⋯ | ⋯ |
| $3d^2(^3F)4d$ | $^4D$ | 5/2 | 45,983.23 | ⋯ | ⋯ | ⋯ | ⋯ | ⋯ | ⋯ | ⋯ |
| $3d^2(^3F)4d$ | $^4H$ | 11/2 | 45,985.91 | ⋯ | ⋯ | ⋯ | ⋯ | ⋯ | ⋯ | ⋯ |
| $3d^2(^3F)4d$ | $^4D$ | 7/2 | 46,042.69 | **1.57**(**37**)[f] | ⋯ | 2(4) | 4,12 | −3.45(0.037) | ⋯ | Van Deelen (2017) |
| $3d^2(^3F)4d$ | $^4H$ | 13/2 | 46,054.28 | ⋯ | ⋯ | 1(3) | 1,2,4 | 1.05(0.0174) | ⋯ | Van Deelen (2017) |

**Notes.**
[a] For the level designation, we used NIST Atomic Spectra Database (ASD) version 5.8 (Kramida et al. 2020), and the energy level values are taken from Ahmed & Verges (1977).
[b] The HFS constants A and B with their uncertainties ΔA and ΔB, respectively. The uncertainties in parentheses are in the last digit of the value. The recommended values of HFS A and B constants are given in bold font.
[c] $1\ mK = 10^{-3}\ cm^{-1}$.
[d] The number of lines used to determine the HFS constants. In parentheses are the total number of measurements of these lines in spectra using Ar and Ne as a buffer gas.
[e] HFS constants $A^{prev}$ and $B^{prev}$ are taken from the reference given in the last column of this table, and the uncertainties in parentheses are in the last digit of the value, except for the measurements by Van Deelen (2017). The recommended HFS A and B constants are given in bold font.
[f] The HFS A constant is revised in this work.

**Table 3**
New and Remeasured Hyperfine Structure Constants for Odd Levels in Sc I

| Level Designation[a] | | | Energy[a] ($cm^{-1}$) | This Work | | | | Previous Experimental Results[e] | | |
|---|---|---|---|---|---|---|---|---|---|---|
| Configuration | Term | J | | A(ΔA)[b] (mK)[c] | B(ΔB)[b] (mK)[c] | No. of Lines[d] | Spectra Used | $A(\Delta A)^{prev}$ (mK)[c] | $B(\Delta B)^{prev}$ (mK)[c] | Reference |
| 3d4s($^3$D)4p | $^4F^*$ | 3/2 | 15,672.58 | **−4.59(1)** | **−0.89(15)** | 4(19) | 1–9,12 | −4.63(2) | ⋯ | Aboussaïd et al. (1996) |
| 3d4s($^3$D)4p | $^4F^*$ | 5/2 | 15,756.57 | **9.66(1)** | ⋯ | 5(17) | 1,2,4–9 | 9.67(10) | ⋯ | Aboussaïd et al. (1996) |
| 3d4s($^3$D)4p | $^4F^*$ | 7/2 | 15,881.75 | **11.99(1)** | **−1.30(18)** | 2(4) | 1,2 | 11.99(1) | ⋯ | Aboussaïd et al. (1996) |
| 3d4s($^3$D)4p | $^4F^*$ | 9/2 | 16,026.62 | **13.04(1)** | **−2.11(46)** | 2(4) | 1,2 | 13.02(1) | ⋯ | Aboussaïd et al. (1996) |
| 3d4s($^3$D)4p | $^4D^*$ | 1/2 | 16,009.77 | **−17.04(6)** | ⋯ | 4(23) | 1–5,8,9,12 | −17.27(5) | ⋯ | Aboussaïd et al. (1996) |
| 3d4s($^3$D)4p | $^4D^*$ | 3/2 | 16,021.82 | **11.42(8)** | ⋯ | 3(8) | 1,2,5–9 | 11.62(4) | ⋯ | Aboussaïd et al. (1996) |
| 3d4s($^3$D)4p | $^4D^*$ | 5/2 | 16,141.06 | **10.21(2)** | ⋯ | 5(22) | 1,2,4–9,12 | 10.22(2) | ⋯ | Aboussaïd et al. (1996) |
| 3d4s($^3$D)4p | $^4D^*$ | 7/2 | 16,210.85 | **15.27(1)** | **0.73(39)** | 3(5) | 1,2 | 15.27(1) | ⋯ | Aboussaïd et al. (1996) |
| 3d4s($^1$D)4p | $^2D^*$ | 5/2 | 16,022.73 | **22.05(6)** | **−15.81(86)** | 3(24) | 1–12 | 21.50(9) | ⋯ | Kopfermann & Rasmussen (1934) |
| 3d4s($^1$D)4p | $^2D^*$ | 3/2 | 16,096.90 | **−12.45(2)** | ⋯ | 4(26) | 1–12 | 12.41(5) | ⋯ | Aboussaïd et al. (1996) |
| 3d4s($^3$D)4p | $^4P^*$ | 1/2 | 18,504.06 | **40.00(15)** | ⋯ | 1(6) | 1,2,4,5,9,12 | 39.93(10) | ⋯ | Başar et al. (2004) |
| 3d4s($^3$D)4p | $^4P^*$ | 3/2 | 18,515.69 | **24.42(6)** | ⋯ | 2(9) | 1–5,7,8,12 | 24.48(10) | ⋯ | Başar et al. (2004) |
| 3d4s($^3$D)4p | $^4P^*$ | 5/2 | 18,571.41 | **19.61(2)** | ⋯ | 5(20) | 1–5,9,12 | 19.68(17) | ⋯ | Başar et al. (2004) |
| 4s$^2$4p | $^2P^*$ | 1/2 | 18,711.02 | **2.96(3)** | ⋯ | 1(8) | 5–12 | 2.96(0.0054) | ⋯ | Van Deelen (2017) |
| 4s$^2$4p | $^2P^*$ | 3/2 | 18,855.74 | **20.38(1)** | **−0.93(7)** | 2(22) | 1–12 | 20.36(0.00033) | ⋯ | Van Deelen (2017) |
| 3d4s($^1$D)4p | $^2F^*$ | 5/2 | 21,032.75 | **−1.33(1)** | **−1.61(19)** | 1(7) | 5–11 | −1.23(53) | ⋯ | Başar et al. (2004) |
| 3d4s($^1$D)4p | $^2F^*$ | 7/2 | 21,085.85 | **12.56(1)** | **−2.63(14)** | 1(10) | 3–12 | 12.38(27) | ⋯ | Başar et al. (2004) |
| 3d4s($^1$D)4p | $^2P^*$ | 1/2 | 24,656.72 | **14.74(16)** | ⋯ | 1(2) | 1,2 | 14.32(0.0082) | ⋯ | Van Deelen (2017) |
| 3d4s($^1$D)4p | $^2P^*$ | 3/2 | 24,656.88 | **0.42(3)**[f] | **−6.36(42)** | 1(2) | 1,2 | −0.94(0.015) | ⋯ | Van Deelen (2017) |
| 3d4s($^3$D)4p | $^2D^*$ | 3/2 | 24,866.17 | **9.27(18)** | ⋯ | 3(7) | 1,2,5 | 8.57(0.024) | −0.57(23) | Van Deelen (2017) |
| 3d4s($^3$D)4p | $^2D^*$ | 5/2 | 25,014.21 | **1.03(12)** | ⋯ | 1(2) | 1,2 | 1.10(17) | ⋯ | Başar et al. (2004) |
| 3d4s($^3$D)4p | $^2F^*$ | 5/2 | 25,584.64 | **5.72(5)** | ⋯ | 3(7) | 1,2,10 | 5.88(01) | ⋯ | Van Deelen (2017) |
| 3d4s($^3$D)4p | $^2F^*$ | 7/2 | 25,724.68 | **4.16(10)** | ⋯ | 1(5) | 1,3,5,10 | 4.26(0.011) | 1.90(25) | Van Deelen (2017) |
| 3d$^2$($^3$F)4p | $^4G^*$ | 5/2 | 29,022.82 | **9.31(4)** | **−0.98(24)** | 1(11) | 2–12 | 9.36(10) | −2(1) | Singh et al. (1991) |
| 3d$^2$($^3$F)4p | $^4G^*$ | 7/2 | 29,096.18 | **4.91(1)** | ⋯ | 1(5) | 5–9 | 5.03(0.0019) | ⋯ | Van Deelen (2017) |
| 3d$^2$($^3$F)4p | $^4G^*$ | 9/2 | 29,189.84 | **2.78(2)** | **−0.83(15)** | 2(23) | 1–12 | ⋯ | ⋯ | ⋯ |
| 3d$^2$($^3$F)4p | $^4G^*$ | 11/2 | 29,303.51 | **1.49(1)** | **−1.31(25)** | 1(7) | 5–11 | 1.84(3) | 0.83(1.00) | Singh et al. (1991) |
| 3d4s($^3$D)4p | $^2P^*$ | 1/2 | 30,573.17 | **8.00(62)** | ⋯ | 2(2) | 1 | 6.63(0.021) | ⋯ | Van Deelen (2017) |
| 3d4s($^3$D)4p | $^2P^*$ | 3/2 | 30,706.66 | **10.92(12)** | ⋯ | 2(3) | 1,2 | 10.26(01) | ⋯ | Van Deelen (2017) |
| 3d$^2$($^3$F)4p | $^4F^*$ | 3/2 | 31,172.70 | **9.35(1)** | ⋯ | 1(12) | 1–12 | ⋯ | ⋯ | ⋯ |
| 3d$^2$($^3$F)4p | $^4F^*$ | 5/2 | 31,215.81 | **4.04(1)** | ⋯ | 2(16) | 4,5,7–12 | ⋯ | ⋯ | ⋯ |
| 3d$^2$($^3$F)4p | $^4F^*$ | 7/2 | 31,275.39 | **2.25(1)** | ⋯ | 2(20) | 1–12 | ⋯ | ⋯ | ⋯ |
| 3d$^2$($^3$F)4p | $^4F^*$ | 9/2 | 31,350.84 | **1.48(1)** | ⋯ | 2(20) | 1–12 | ⋯ | ⋯ | ⋯ |
| 3d$^2$($^3$F)4p | $^4D^*$ | 1/2 | 32,637.40 | **14.44(3)** | ⋯ | 1(10) | 3–12 | 16.8236(20) | ⋯ | Van Deelen (2017) |
| 3d$^2$($^3$F)4p | $^4D^*$ | 3/2 | 32,659.30 | **4.21(5)** | ⋯ | 2(18) | 3–12 | ⋯ | ⋯ | ⋯ |
| 3d$^2$($^3$F)4p | $^4D^*$ | 5/2 | 32,696.84 | **2.35(3)** | ⋯ | 1(7) | 4,6–10,12 | ⋯ | ⋯ | ⋯ |
| 3d$^2$($^3$F)4p | $^4D^*$ | 7/2 | 32,751.50 | **1.40(1)**[f] | ⋯ | 2(18) | 1,2 | −2.66(0.0019) | ⋯ | Van Deelen (2017) |
| 3d$^2$($^3$F)4p | $^2G^*$ | 7/2 | 33,055.98 | **5.65(4)** | ⋯ | 4(9) | 1,2,4,9 | 5.62(0.0098) | ⋯ | Van Deelen (2017) |
| 3d$^2$($^3$F)4p | $^2G^*$ | 9/2 | 33,151.20 | **2.36(13)** | ⋯ | 1(3) | 1,2,9 | 2.36(0.014) | ⋯ | Van Deelen (2017) |
| 3d$^2$($^3$F)4p | $^2F^*$ | 5/2 | 33,153.79 | **7.57(6)** | ⋯ | 1(4) | 1–3,10 | 7.64(0.0059) | ⋯ | Van Deelen (2017) |
| 3d$^2$($^3$F)4p | $^2F^*$ | 7/2 | 33,278.40 | **1.33(2)**[f] | ⋯ | 3(9) | 1–3,5,9 | −2.74(13) | ⋯ | Singh et al. (1991) |
| 3d$^2$($^3$F)4p | $^2D^*$ | 3/2 | 33,614.88 | **7.86(11)** | ⋯ | 3(15) | 1–12 | 7.51(27) | ⋯ | Başar et al. (2004) |
| 3d$^2$($^3$F)4p | $^2D^*$ | 5/2 | 33,707.06 | **1.39(5)** | ⋯ | 4(13) | 1–5,7,9 | 1.47(27) | ⋯ | Başar et al. (2004) |
| 3d$^2$($^3$P)4p | $^2S^*$ | 1/2 | 35,346.35 | **−10.59(35)** | ⋯ | 2(4) | 1,2 | −10.43(0.017) | ⋯ | Van Deelen (2017) |
| 3d$^2$($^1$D)4p | $^2F^*$ | 5/2 | 36,666.42 | **4.96(6)** | ⋯ | 3(20) | 1–5,7–12 | ⋯ | ⋯ | ⋯ |

**Table 3**
(Continued)

| Level Designation[a] | | | Energy[a] ($cm^{-1}$) | This Work | | | | Previous Experimental Results[e] | | |
|---|---|---|---|---|---|---|---|---|---|---|
| Configuration | Term | J | | A(ΔA)[b] (mK)[c] | B(ΔB)[b] (mK)[c] | No. of Lines[d] | Spectra Used | $A(\Delta A)^{prev}$ (mK)[c] | $B(\Delta B)^{prev}$ (mK)[c] | Reference |
| $3d^2(^1D)4p$ | $^2F^*$ | 7/2 | 36,730.12 | **3.20(2)** | … | 1(11) | 1–12 | … | … | … |
| $3d^2(^3P)4p$ | $^4D^*$ | 1/2 | 36,764.20 | **11.97(76)** | … | 1(11) | 1–5,7–12 | … | … | … |
| $3d^2(^3P)4p$ | $^4D^*$ | 3/2 | 36,793.65 | **1.86(15)** | … | 2(3) | 4,12 | … | … | … |
| $3d^2(^3P)4p$ | $^4D^*$ | 5/2 | 36,860.20 | **0.98(14)** | … | 1(2) | 4,12 | … | … | … |
| $3d^2(^3P)4p$ | $^4D^*$ | 7/2 | 36,959.03 | **0.75(12)** | … | 1(12) | 1–12 | … | … | … |
| $3d^2(^1D)4p$ | $^2D^*$ | 3/2 | 36,933.91 | **5.41(5)** | … | 2(16) | 3–5,8–12 | … | … | … |
| $3d^2(^1D)4p$ | $^2D^*$ | 5/2 | 37,039.57 | **4.44(3)** | … | 2(13) | 3–5,7–12 | … | … | … |
| $3d^2(^1D)4p$ | $^2P^*$ | 3/2 | 37,086.02 | **7.19(3)** | … | 2(18) | 3–5,7–12 | … | … | … |
| $3d^2(^1D)4p$ | $^2P^*$ | 1/2 | 37,125.40 | **3.68(5)** | … | 1(8) | 4,5,7–12 | … | … | … |
| $3d^2(^3P)4p$ | $^4S^*$ | 3/2 | 37,486.86 | **−4.65(12)** | … | 2(17) | 3–5,7–12 | … | … | … |
| $3d^2(^3P)4p$ | $^4P^*$ | 1/2 | 37,877.78 | **−1.74(25)** | … | 1(7) | 4,9,11,12 | … | … | … |
| $3d^2(^3P)4p$ | $^4P^*$ | 3/2 | 37,908.50 | **1.48(7)** | … | 3(21) | 3–5,7–12 | … | … | … |
| $3d^2(^3P)4p$ | $^4P^*$ | 5/2 | 37,964.89 | **1.59(30)** | … | 1(2) | 4,12 | … | … | … |
| $3d^2(^1G)4p$ | $^2H^*$ | 9/2 | 39,153.14 | **4.66(13)** | … | 1(12) | 1–12 | … | … | … |
| $3d^2(^1G)4p$ | $^2H^*$ | 11/2 | 39,248.82 | **3.66(13)** | … | 1(12) | 1–12 | … | … | … |
| $3d^2(^1G)4p$ | $^2G^*$ | 7/2 | 39,392.79 | **4.31(11)** | … | 1(12) | 1–12 | … | … | … |
| $3d^2(^1G)4p$ | $^2G^*$ | 9/2 | 39,423.39 | **3.71(9)** | … | 2(15) | 1–12 | … | … | … |
| $3d4s(^3D)5p$ | $^4F^*$ | 3/2 | 39,949.75 | **−12.81(25)** | … | 2(3) | 1,2 | −11.03(0.011) | −1.30 | Van Deelen (2017) |
| $3d4s(^3D)5p$ | $^4F^*$ | 5/2 | 39,989.58 | **3.93(14)** | … | 1(2) | 1,2 | 3.87(0.0034) | −0.17(0.045) | Van Deelen (2017) |
| $3d4s(^3D)5p$ | $^4F^*$ | 7/2 | 40,048.72 | **10.69(3)** | … | 1(2) | 1,2 | 10.62(0.0029) | … | Van Deelen (2017) |
| $3d4s(^3D)5p$ | $^4F^*$ | 9/2 | 40,145.90 | **14.37(4)** | … | 1(2) | 1,2 | 14.39(0.0034) | −2.96(12) | Van Deelen (2017) |
| $3d4s(^3D)5p$ | $^4D^*$ | 1/2 | 40,044.63 | **−23.63(24)** | … | 2(4) | 1,2 | −24.52(0.011) | … | Van Deelen (2017) |
| $3d4s(^3D)5p$ | $^4D^*$ | 3/2 | 40,073.49 | **11.81(12)** | … | 3(5) | 1,2 | 11.70(0.011) | −0.82(0.04) | Van Deelen (2017) |
| $3d4s(^3D)5p$ | $^4D^*$ | 5/2 | 40,128.13 | **13.57(7)** | … | 4(6) | 1,2,4,12 | 13.95(0.011) | … | Van Deelen (2017) |
| $3d4s(^3D)5p$ | $^4D^*$ | 7/2 | 40,210.88 | **16.79(5)** | … | 1(2) | 1,2 | 16.86(0.0021) | 1.02(0.059) | Van Deelen (2017) |
| $3d4s(^3D)5p$ | $^2F^*$ | 5/2 | 40,104.19 | **3.41(11)** | … | 1(5) | 4,5,9,10,12 | 3.34(0.0037) | −1.13(0.044) | Van Deelen (2017) |
| $3d4s(^3D)5p$ | $^2F^*$ | 7/2 | 40,151.08 | **14.79(22)** | … | 2(2) | 1,4 | 14.82(0.0053) | −3.35(15) | Van Deelen (2017) |
| $3d4s(^3D)5p$ | $^2D^*$ | 3/2 | 40,347.34 | … | … | … | … | … | … | … |
| $3d4s(^3D)5p$ | $^2D^*$ | 5/2 | 40,351.30 | **9.49(4)** | … | 3(10) | 1,3,4,5,9,10,12 | … | … | … |
| $3d4s(^3D)5p$ | $^2P^*$ | 1/2 | 40,499.71 | … | … | … | … | … | … | … |
| $3d4s(^3D)5p$ | $^2P^*$ | 3/2 | 40,594.07 | **4.54(12)** | … | 1(3) | 4,9,12 | … | … | … |
| $3d4s(^3D)5p$ | $^4P^*$ | 1/2 | 40,595.28 | **56.82(30)** | … | 2(4) | 1,2 | 56.94(0.032) | … | Van Deelen (2017) |
| $3d4s(^3D)5p$ | $^4P^*$ | 3/2 | 40,644.64 | **32.07(13)** | … | 3(4) | 1,2,4 | 30.90(0.016) | 0.56(14) | Van Deelen (2017) |
| $3d4s(^3D)5p$ | $^4P^*$ | 5/2 | 40,715.42 | **22.22(8)** | … | 2(3) | 1,2 | 22.15(0.012) | −1.52(17) | Van Deelen (2017) |
| $3d^2(^3P)4p$ | $^2D^*$ | 3/2 | 41,153.42 | **−6.51(14)** | … | 4(7) | 1,4,9,12 | −6.70(0.0054) | 1.10(11) | Van Deelen (2017) |
| $3d^2(^3P)4p$ | $^2D^*$ | 5/2 | 41,162.52 | **15.65(6)** | … | 4(8) | 1,2,4,9,12 | 15.67(0.0043) | −0.72(17) | Van Deelen (2017) |
| $3d4s(^1D)5p$ | $^2P^*$ | 3/2 | 42,780.41 | **25.29(29)** | … | 3(4) | 1,2,4 | 25.93(0.028) | −7.79(22) | Van Deelen (2017) |
| $3d4s(^1D)5p$ | $^2P^*$ | 1/2 | 42,819.49 | … | … | … | … | … | … | … |
| $3d4s(^1D)5p$ | $^2F^*$ | 5/2 | 42,938.79 | … | … | … | … | … | … | … |
| $3d4s(^1D)5p$ | $^2F^*$ | 7/2 | 42,978.81 | … | … | … | … | 2.24(90) | … | Van Deelen (2017) |
| $3d4s(^1D)5p$ | $^2D^*$ | 3/2 | 43,170.45 | **11.42(12)** | … | 1(1) | 1 | 11.08(0.014) | −1.22(15) | Van Deelen (2017) |
| $3d4s(^1D)5p$ | $^2D^*$ | 5/2 | 43,252.56 | **7.32(37)** | … | 1(2) | 1,12 | … | … | … |
| $3d^2(^1G)4p$ | $^2F^*$ | 5/2 | 43,830.12 | **0.38(17)** | … | 2(3) | 5, 8 | … | … | … |
| $3d^2(^1G)4p$ | $^2F^*$ | 7/2 | 43,860.12 | **8.39(11)** | … | 2(8) | 1–4,9–12 | 7.77(39) | … | Van Deelen (2017) |

**Table 3**
(Continued)

| Level Designation[a] | | | Energy[a] ($cm^{-1}$) | This Work | | | | Previous Experimental Results[e] | | |
|---|---|---|---|---|---|---|---|---|---|---|
| Configuration | Term | J | | A(ΔA)[b] (mK)[c] | B(ΔB)[b] (mK)[c] | No. of Lines[d] | Spectra Used | $A(\Delta A)^{prev}$ (mK)[c] | $B(\Delta B)^{prev}$ (mK)[c] | Reference |
| $3d^2(^3P)4p$ | $^2P^*$ | 1/2 | 44,105.45 | **15.43(93)** | … | 3(5) | 1, 4,12 | … | … | … |
| $3d^2(^3P)4p$ | $^2P^*$ | 3/2 | 44,189.29 | … | … | … | … | … | … | … |

**Notes.**
[a] For the level designation, we used NIST ASD version 5.8 (Kramida et al. 2020), and the energy level values are taken from Ahmed & Verges (1977).
[b] The HFS constants A and B with their uncertainties ΔA and ΔB, respectively. The uncertainties in the parentheses are in the last digit of the value. The recommended values of HFS A and B constants are given in bold font.
[c] 1 mK = $10^{-3}$ $cm^{-1}$.
[d] The number of lines used to determine the HFS constants. In parentheses are the total number of measurements of these lines in spectra using Ar and Ne as a buffer gas.
[e] HFS constants $A^{prev}$ and $B^{prev}$ are taken from the reference given in the last column of this table, and the uncertainties in parentheses are in the last digit of the value, except for the measurements by Van Deelen (2017). The recommended HFS A and B constants are given in bold font.
[f] The HFS A constant is revised in this work.

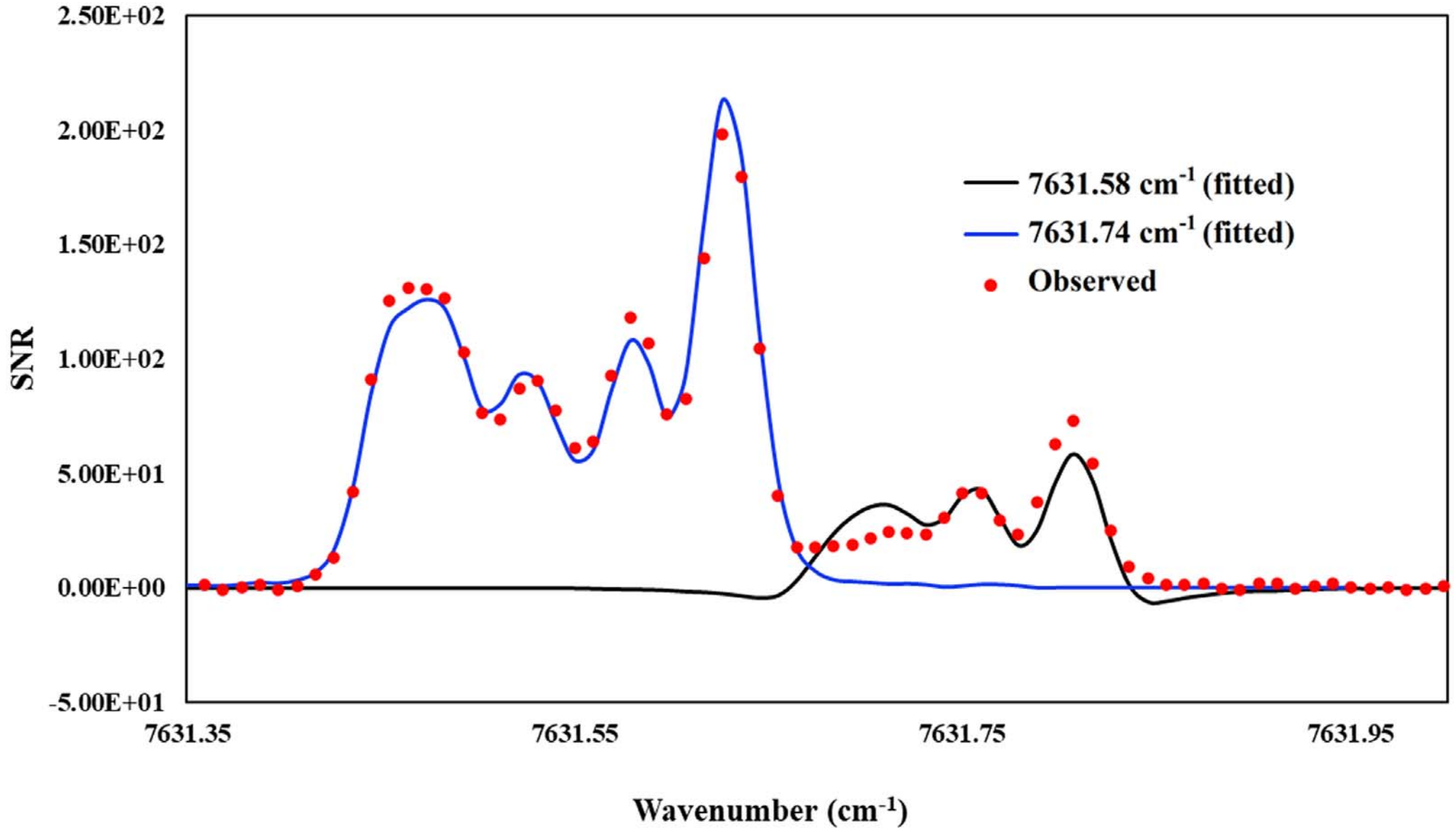

**Figure 2.** Section of the FT spectrum of Sc I showing the fitted line profiles of the $3d^2(^1D)4s\ ^2D_{3/2}$–$3d4s(^1D)4p\ ^2P^*_{3/2}$ (76,31.58 cm$^{-1}$) and $3d^2(^1D)4s\ ^2D_{3/2}$–$3d4s(^1D)4p\ ^2P^*_{3/2}$ (7631.74 cm$^{-1}$) transitions in spectrum 1 (Sc051815.002).

**Table 4**
Hyperfine Structure (HFS) Line Component Patterns for Transitions of Sc I Computed from Known HFS Coefficients

| Transition Wavenumber[a] (cm$^{-1}$) | $F_{upp}$ | $F_{low}$ | Component Position (cm$^{-1}$) | Strength Normalized to 1.0 for Each Transition |
|---|---|---|---|---|
| 39,905.2 | 5 | 6 | 0.030152 | 0.270833 |
| 39,905.2 | 5 | 5 | 0.051992 | 0.064167 |
| 39,905.2 | 4 | 5 | −0.007060 | 0.165000 |
| 39,905.2 | 5 | 4 | 0.070192 | 0.008750 |
| 39,905.2 | 4 | 4 | 0.011142 | 0.092812 |
| 39,905.2 | 3 | 4 | −0.036100 | 0.085937 |
| 39,905.2 | 4 | 3 | 0.025702 | 0.023438 |
| 39,905.2 | 3 | 3 | −0.021540 | 0.091146 |
| 39,905.2 | 2 | 3 | −0.056970 | 0.031250 |
| 39,905.2 | 3 | 2 | −0.010620 | 0.041667 |
| 39,905.2 | 2 | 2 | −0.046050 | 0.062500 |
| 39,905.2 | 2 | 1 | −0.038770 | 0.062500 |
| 39,935.9 | 6 | 6 | −0.002010 | 0.225694 |
| 39,935.9 | 5 | 6 | −0.022470 | 0.045139 |
| 39,935.9 | 6 | 5 | 0.019827 | 0.045139 |
| 39,935.9 | 5 | 5 | −0.000630 | 0.115456 |
| 39,935.9 | 4 | 5 | −0.017680 | 0.068571 |
| 39,935.9 | 5 | 4 | 0.017567 | 0.068571 |
| 39,935.9 | 4 | 4 | 0.000517 | 0.045268 |
| 39,935.9 | 3 | 4 | −0.013120 | 0.073661 |
| 39,935.9 | 4 | 3 | 0.015078 | 0.073661 |
| 39,935.9 | 3 | 3 | 0.001438 | 0.008681 |
| 39,935.9 | 2 | 3 | −0.008790 | 0.063492 |
| 39,935.9 | 3 | 2 | 0.012358 | 0.063492 |
| 39,935.9 | 2 | 2 | 0.002127 | 0.000496 |
| 39,935.9 | 1 | 2 | −0.004690 | 0.040179 |
| 39,935.9 | 2 | 1 | 0.009407 | 0.040179 |
| 39,935.9 | 1 | 1 | 0.002587 | 0.022321 |

**Notes.** Center-of-gravity wavenumbers are given with component positions relative to those values. This table is available in its entirety in machine-readable form.
[a] Wavenumber is calculated from the experimental energy level values (Ahmed & Verges 1977) given in Tables 2 and 3.

(This table is available in its entirety in machine-readable form.)

configurations, improving the accuracy of the previously known HFS constants. Mansour et al. (1989a) and Mansour et al. (1989b) improved the accuracy of the HFS A and B constants for the $3d^2$ ($^1D_2$, $^3P_{1,2}$, $^1G_4$) levels by factors of 150 and 300, respectively. Also, the accuracy of the HFS A and B constants for the $3d4p(^3D^*_{1,2,3}$, $^3P^*_2$, $^1F^*_3)$ levels were improved by at least a factor of five, and new HFS A constants were measured for the $3d^2\ ^3F_{2,3,4}$ and $3d4p\ ^3F^*_4$ levels. Further, Beck (1992) did the calculation using the nonrelativistic configuration interaction Hartree–Fock method, and found good agreement with the experimental HFS constants. The most recent experimental work was done by Villemoes et al. (1992), using the collinear fast-ion beam laser spectroscopy technique to study 12 levels in the 3d4s, $3d^2$, and 3d4p configurations. Of all these works, the most recent and accurate data come from Villemoes et al. (1992), Mansour et al. (1989a), and Mansour et al. (1989b), with the latter employing the ultrahigh-resolution ABMR technique.

All the experimental HFS A and B constant from the five experimental papers agree with each other within the experimental uncertainty. We used the previously known values of the HFS A and B constants (given in columns 9 and 10 of Table 5) as fixed parameters in the fitting program Xgremlin (Nave et al. 2015) to measure the HFS A constants for the $3d4p\ ^1P^*_1$, $4s4p\ ^3P^*_{1,2}$ and $3d4d\ ^3G_{3,4,5}$ levels. In this work, we measured the HFS A constant for three high odd-parity levels and the three even-parity levels required for Lawler et al. (2019) to determine the Sc II abundance in MP stars. These HFS constants are given in Table 5.

## 5. Uncertainty Estimation

The uncertainty in the HFS constants depends on a number of factors. The principal component of the uncertainty is the statistical uncertainty in the calculated position of the HFS transition components, which is calculated in the Xgremlin fitting program (Nave et al. 2015). For weak transitions with S/N < 10, the wavenumber uncertainty can be large because some component lines may fall below the noise level. The ratio

**Table 5**
Hyperfine Structure Constants for Sc II

| Level Designation[a] | | | Energy[a] ($cm^{-1}$) | This Work | | | | Previous Experimental Results[e] | | |
|---|---|---|---|---|---|---|---|---|---|---|
| Configuration | Term | J | | A(ΔA)[b] (mK)[c] | B(ΔB)[b] (mK)[c] | No. of Line[d] | Spectra Used | $A(\Delta A)^{prev}$ (mK)[c] | $B(\Delta B)^{prev}$ (mK)[c] | Reference |
| 3d4s | $^3D$ | 1 | 0 | | | | | −16.01 (7) | −0.43(10) | Villemoes et al. (1992) |
| 3d4s | $^3D$ | 2 | 67.72 | | | | | 17.01(3) | −1.0(4) | Villemoes et al. (1992) |
| 3d4s | $^3D$ | 3 | 177.76 | | | | | 21.84(2) | −2.10(77) | Villemoes et al. (1992) |
| 3d4s | $^1D$ | 2 | 2540.95 | | | | | 4.28(3) | −1.30(37) | Villemoes et al. (1992) |
| $3d^2$ | $^3F$ | 2 | 4802.87 | | | | | 9.6957(4) | −0.3516(28) | Mansour et al. (1989b) |
| $3d^2$ | $^3F$ | 3 | 4883.57 | | | | | 3.7918(1) | −0.4208(13) | Mansour et al. (1989b) |
| $3d^2$ | $^3F$ | 4 | 4987.79 | | | | | 1.2795(1) | −0.549(3) | Mansour et al. (1989b) |
| $3d^2$ | $^1D$ | 2 | 10,944.56 | | | | | 4.9821(1) | 0.26078(2) | Mansour et al. (1989b) |
| $3d^2$ | $^3P$ | 1 | 12,101.50 | | | | | −3.58584(7) | −0.41018(2) | Mansour et al. (1989b) |
| $3d^2$ | $^3P$ | 2 | 12,154.42 | | | | | −0.92503(7) | 0.7381(8) | Mansour et al. (1989b) |
| $3d^2$ | $^1G$ | 4 | 14,261.32 | | | | | 4.51085(3) | −2.1161(13) | Mansour et al. (1989b) |
| 3d4p | $^1D^*$ | 2 | 26,081.34 | | | | | 7.19(3) | 0.60(23) | Arnesen et al. (1982) |
| 3d4p | $^3F^*$ | 2 | 27,443.71 | | | | | 12.235(3) | −1.33(5) | Mansour et al. (1989b) |
| 3d4p | $^3F^*$ | 3 | 27,602.45 | | | | | 6.85(2) | −2.33(6) | Mansour et al. (1989b) |
| 3d4p | $^3F^*$ | 4 | 27,841.35 | | | | | 3.412(3) | −2.802(7) | Mansour et al. (1989b) |
| 3d4p | $^3D^*$ | 1 | 27,917.78 | | | | | 10.164(7) | 0.15(3) | Mansour et al. (1989b) |
| 3d4p | $^3D^*$ | 2 | 28,021.29 | | | | | 4.180(3) | 0.334(3) | Mansour et al. (1989b) |
| 3d4p | $^3D^*$ | 3 | 28,161.17 | | | | | 3.3190(3) | 0.701(7) | Mansour et al. (1989b) |
| 3d4p | $^3P^*$ | 1 | 29,742.16 | | | | | 8.506(7) | 0.33(3) | Mansour et al. (1989b) |
| 3d4p | $^3P^*$ | 2 | 29,823.93 | | | | | 3.542(3) | 0.667(3) | Mansour et al. (1989b) |
| 3d4p | $^1P^*$ | 1 | 30,815.70 | 6.31(2.00) | … | 2(4) | 3,4,10 | … | … | … |
| 3d4p | $^1F^*$ | 3 | 32,349.98 | | | | | 6.37(1) | −2.74(1) | Mansour et al. (1989b) |
| 4s4p | $^3P^*$ | 2 | 39,115.04 | 36.10(0.71) | … | 2(8) | 3,4,10,12 | … | … | … |
| 4s4p | $^3P^*$ | 1 | 39,345.52 | 44.83(0.37) | … | 2(8) | 3,4,10,12 | … | … | … |
| 3d4d | $^3G$ | 3 | 60,267.16 | 4.58(0.39) | … | 1(2) | 4,12 | … | … | … |
| 3d4d | $^3G$ | 4 | 60,348.46 | 2.89(1.22) | … | 1(2) | 4,12 | … | … | … |
| 3d4d | $^3G$ | 5 | 60,457.12 | 3.48(2.50) | … | 2(4) | 4,12 | … | … | … |

**Note.**
[a] The energy level values and level designations are taken from NIST ASD version 5.8 (Kramida et al. 2020).
[b] The HFS constants A and B with their uncertainties ΔA and ΔB, respectively. The uncertainties in the parentheses are in the last digit of the value.
[c] $1\ mK = 10^{-3}\ cm^{-1}$
[d] The number of lines used to determine the HFS constants. In parentheses are the total number of measurements of these lines in spectra using Ar and Ne as a buffer gas.
[e] HFS constants $A^{prev}$ and $B^{prev}$ are taken from the reference given in the last column of this table, and the uncertainties in parentheses are in the last digit of the values.

between the HFS line width and the HFS line separation is also an important factor in determining the uncertainty. A line profile with resolved components has lower uncertainties than those for a profile with blended components or features from another transition. We used several different transitions in multiple spectra to measure the HFS constants for a single level ($E_J$), and the final value is a weighted average of the values determined from the line profile fits using a two-step process (Townly-Smith et al. 2016). First, we derive the weighted average of the HFS A constants ($A_{\rm av}$) for each transition being measured in each spectrum $i$ using:

$$A_{av} = \sum_{i}^{n} A_i \delta_i^{-2} \Big/ \sum_{i}^{n} \delta_i^{-2}, \quad (4)$$

where $A_i$ is the A constants determined from the line in the $i$th spectrum, $\delta_i$ is its standard deviation obtained from a least-square fit of the line, and $n$ is the number of spectra used for each spectral line. Weight is defined as the inverse square of the standard deviation $\delta_i$. After deriving the weighted average for each line from multiple spectra, we estimated one standard uncertainty for each line, $\Delta A_{av}$, using the greater of:

$$\Delta A_{av} = \sqrt{\sum_{i}^{n} [(A_i - A_{av})^2 \delta_i^{-2}] / [(n-1) \sum_{i}^{n} \delta_i^{-2}]} \quad (5)$$

and

$$\sqrt{1 / \sum_{i}^{n} \delta_i^{-2}}. \quad (6)$$

Equation (6) prevents a misleading low uncertainty that can arise from a small number of measured HFS A constants being in coincidentally good agreement. Using this procedure takes into account both the measured standard deviations and the actual distribution of the measured HFS A constant from Equation (5). Next, we added the uncertainty of the HFS A constant of the other level involved in the transition in quadrature to the result from Equations (5) and (6) to give the uncertainty determined from all measurements using that particular transition.

In the second step, we again used Equation (4) to determine the weighted average of the HFS A constants ($A_{\rm av}$), and estimated one standard uncertainty $\Delta A$ for level ($E_{\rm j}$) using the greater of Equations (5) or (6), but this time summing over all transitions, $i$, that can be used to determine the HFS A constant of the level ($E_{\rm j}$), with $n$ now being the number of transitions. The HFS A constant and its one standard uncertainty ($\Delta A$) are given in Tables 2, 3, and 5. For the HFS A constant and its uncertainty, a detailed procedure with an example can be found in Townly-Smith et al. (2016).

## 6. Summary

In Sc I, we have measured HFS dipole A constants for 176 levels using NIST FT spectrometers in the wavelength region of 200–2500 nm. In this work, we have measured the HFS A constants for 91 even-parity configurations ($3d^24s$, 3d4s4d, 3d4s5s, $3d^3$, $3d4p^2$, $3d^25s$, 3d4s6s, $4s^24d$) and 85 levels of odd-parity configurations (3d4s4p, $3d^24p$, $4s^24p$, 3d4s5p). We have also measured the HFS quadrupole B constant for 4 even-parity and 12 odd-parity levels. 90% of the known levels of the 3d4s (4d + 5s + 6s + 4p + 5p), $3d^3$, $3d4p^2$, $3d^2$(4s + 5s + 4p) and $4s^2$(4d + 4p) configurations in Sc I now have measured HFS dipole A constants from this study.

For Sc II, we measured the HFS A constant for six levels of 3d4p, 3d4d, and 4s4p configurations.

We thank James E. Lawler for many helpful discussions and suggestions. NASA grants NNH17AE08I and NNH14A4781 support this research.

## ORCID iDs

Hala https://orcid.org/0000-0002-3083-4917
G. Nave https://orcid.org/0000-0002-1718-9650